\documentstyle[aps,graphicx,epsf]{revtex}

\begin{document}
\title{On the gravitational instability of a dissipative medium.}
\author{L.S. Garc\'{\i }a Col\'{\i }n $^{ab}$. and \ Alfredo Sandoval-Villalbazo $^{\ast
\dag }$ .}
\address{$^*$Departamento de Ciencias, Universidad\\
Iberoamericana, Col. Lomas de Santa Fe~01210, Mexico~D.F., Mexico}
\address{$^\dag$Relativity and Cosmology Group, School of\\
Computer Science and Mathematics, Portsmouth University,\\
Portsmouth~PO1~2EG, Britain. E-Mail: alfredo.sandoval@port.ac.uk}
\address{$^a$ Departamento de F\'{\i }sica,\\
Universidad Aut\'onoma Metropolitana-Iztapalapa,\\
Av. Pur\'{\i }sima y Michoac\'{a}n S/N.\\
Mexico~D.F., 09340 Mexico. e-mail: lgcs@xanum.uam.mx}
\address{$^b$ El Colegio Nacional, Luis Gonz\'alez Obreg\'on 23, Centro\\
Hist\'orico 06020. M\'exico D.F. M\'exico.}
\maketitle

\begin{abstract}
This paper shows that the ordinary Jeans wave number can be obtained as a
limiting case of a more general approach that includes dissipative effects.
Corrections to the Jeans critical mass associated to viscosity are
established. Some possible implications of the results are finally discussed.
\end{abstract}

\smallskip

\section{Introduction}

Structure formation in fluid systems is closely related to the existence of
growing modes while considering density fluctuations in the hydrodynamical
regime. Relevant growing modes at very small wave numbers have long ago been
identified in gravitational systems, leading to key concepts in present
astrophysics such as the {\em Jeans wave number }and the {\em Jeans mass}
\cite{Jeans}. These parameters correspond to the conditions that a density
fluctuation must satisfy in order to grow up to the point of forming a
structure.

Entropy production plays an important role in realistic
descriptions of thermodynamical processes, and some discussion of
its physical sources may yield interesting results in the analysis
of the time evolution of statistical fluctuations in astrophysical
systems. This work is mainly devoted to the study of the
dispersion relation that arises from the fluctuation analysis of
the thermodynamical variables of a simple dissipative fluid {\em
under the influence of a  gravitational field}. The analysis is
reviewed in the case of non-vanishing viscosity and different
features of the fluid are established. In particular, this work
shows the possibility of viscosity-affected growing modes for some
astrophysical fluids, which may be of interest for both
astrophysicists and cosmologists.

This paper is divided as follows: section two is dedicated to the
hydrodynamical model analysis and the establishment of the
dispersion relation in which the main results are based. Section
three is devoted to the mathematical treatment needed for its
solution. The Jeans instability is recovered in the presence of
viscosity for typical situations present in protogalaxies.
Finally, a discussion of the results obtained and an outline of
future work are included in section four of the paper.

\section{Hydrodynamic model}

In order to establish the basic equations that lie at the core of this work,
we remind the reader that the so-called Navier-Stokes-Fourier equations of
hydrodynamics for a simple fluid arise from the structure of what is now
called Linear Irreversible Thermodynamics (LIT) by supplementing the two
conservation equations for mass and momentum, respectively and the balance
equation for the internal energy, with additional information. Indeed, these
equations read \cite{Berne}.

\begin{equation}
\frac{\partial \rho }{\partial t}+\frac{\partial }{\partial x^{i}}(\rho
\,u^{i})=0  \label{Nav1}
\end{equation}
\begin{equation}
\rho \frac{D\,\,u^{i}}{Dt}+\frac{\partial \Xi ^{\,i\,j}}{\partial
x^{j}} =F^{i}  \label{Nav2}
\end{equation}
\begin{equation}
\rho \frac{D\,\,\varepsilon }{Dt}+\frac{\partial Q^{\,j}}{\partial
x^{j}} =-\Xi ^{\,i\,j}\,\frac{\partial u_{i}}{\partial x^{j}}
\label{Nav3}
\end{equation}
where $\frac{D\,}{Dt}=\frac{\partial }{\partial
\,t}+u^{i}\frac{\partial }{\partial x^{i}}.$

\bigskip Here, $\rho (x^{j},t),\,u^{i}(x^{j},t)$ and $\varepsilon (x^{j},t)$
are the local density, velocity and internal energy, respectively, $\Xi
^{\,i\,j}$ the momentum current (or stress tensor) and $Q^{\,j}$ the heat
flux. All indices run from 1 to 3. In general, for isotropic fluids, $\Xi
^{\,i\,j}=p\delta ^{\,i\,j}+\tau ^{\,i\,j}$, where $p$ is the local
hydrostatic pressure and $\tau ^{\,i\,j}$ the viscous tensor ($\delta
^{\,i\,j}$ is the unitary dyadic). Notice that Eqs. (\ref{Nav1}-\ref{Nav3})
contain fifteen unknowns (including the pressure) and there are only five
equations, so the system is not well determined. If we arbitrarily choose to
describe the states of the fluid through the set of variables $\rho
(x^{j},t),\,u^{i}(x^{j},t),$ $T(x^{j},t),$ where $T$ is the local
temperature, we need nine dynamic equations of state (or constitutive
equations) relating $\Xi ^{\,i\,j}$ and $Q^{\,j}$ to the state variables
plus two local equations of state $p=p(\rho ,T)$ and $\varepsilon
=\varepsilon (\rho ,T).$ If according to the tenets of LIT we choose the
so-called linear constitutive laws, namely
\begin{equation}
\tau ^{\,i\,j}=-2\eta \,\sigma ^{\,i\,j}-\zeta \theta \delta ^{\,i\,j}
\label{DSE1}
\end{equation}
\begin{equation}
Q^{\,j}=-\kappa \delta ^{\,ji}\frac{\partial T}{\partial x^{i}}  \label{DSE3}
\end{equation}
with $\eta $ and $\zeta $, the shear and bulk viscosities,
respectively, $ \sigma ^{\,i\,j}$ the symmetrical traceless part
of the velocity gradient,  $ \theta =\frac{\partial
u^{i}}{\partial x^{i}}$ and $\kappa $ being the thermal
conductivity.

Eqs. (\ref{DSE1}) and (\ref{DSE3}) are the well known constitutive
equations of Navier-Stokes and Fourier, respectively. Substitution
of these equations into Eqs. (\ref{Nav1}-\ref{Nav3}) yields a set
of second order in space, first order in time non-linear coupled
set of partial differential equations for the chosen variables
$\rho ,\,u^{i}$ and $T$. This set, which the reader may seek in
the literature \cite{Groot1}, \cite{GC}, is the so called
Navier-Stokes-Fourier system of hydrodynamic equations.The
non-linearities appearing in such equations have two sources, the
inertial terms $\,u^{i} \frac{\partial }{\partial x^{i}}$ \
arising from the hydrodynamic time derivatives, plus quadratic
terms in the gradients of velocity arising from Eqs. (\ref{DSE1})
and (\ref{DSE3}). Moreover, it should be mentioned that this set
of equations is consistent with the second law of thermodynamics,
the Clausius uncompensated heat , or entropy production, is
strictly positive definite.

\bigskip

Nevertheless, for the purpose of this paper, this set of equations is too
complicated. In order to deal with fluctuations around the equilibrium
state, one assures that for any of two state variables, call them $X(x^{j},t)
$ one can write that,
\begin{equation}
X(x^{j},t)=X_{o}+\delta X(x^{j},t)  \label{fluctuations1}
\end{equation}
where $X_{o}$ is the equilibrium value of $X$ and $\delta X$ the
corresponding fluctuation. Neglecting all terms of order $(\delta
X(x^{j},t))^{2}$ and higher in the NSF non-linear set one finds
the {\em linearized NSF equations of hydrodynamics}
\cite{GC}-\cite{Berne},
\begin{equation}
\frac{\partial }{\partial t}\left( \delta \rho \right) +\rho \,\theta =0
\label{Fluc1}
\end{equation}
\begin{equation}
\rho _{o}\frac{\partial u_{k}}{\partial t}=-\,\frac{1}{\rho
_{o}\,\kappa _{T} }\frac{\partial }{\partial x^{k}}\left( \delta
\rho \right) -\frac{\beta }{ \kappa _{T}}\frac{\partial }{\partial
x^{k}}\left( \delta T\right) +2\eta \delta ^{i\,j}\frac{\partial
u_{k}}{\partial x^{i}\partial x^{j}}-(\frac{2}{3 }\eta -\varsigma
)\frac{\partial }{\partial x^{k}}(\theta )+F_{k} \label{Fluc2}
\end{equation}
\begin{equation}
\frac{\partial }{\partial t}\left( \delta T\right) =D_{T}\,\delta
^{i\,j} \frac{\partial \left( \delta T\right) }{\partial
x^{i}\partial x^{j}}-\frac{ \beta \,T_{o}}{\rho _{o}C_{p}\kappa
_{T}}\theta   \label{Fluc3}
\end{equation}
since $u_{k\;o}=0$, $u_{k}=\delta u_{k}$ and $\frac{\beta }{\kappa
_{T}}=( \frac{\partial p_{0}}{\partial T_{o}})_{\rho _{o}}$,
$\kappa _{T}=\rho _{o}( \frac{\partial p_{0}}{\partial \rho
_{o}})_{T_{o}}$ and $D_{T}=\frac{k}{\rho _{o}C_{v}}$ is the
thermal diffusivity. $C_{p}$ and $C_{v}$ are the specific heats at
constant pressure and constant volume, respectively.

Note now that from Eq. (\ref{Fluc2}), $u_{k\;Trans}\equiv Rot\left(
u_{k}\right) $ uncouples from the hydrodynamic modes, whence, one formally
arrives at the desired set, namely:
\begin{equation}
\frac{\partial }{\partial t}\left( \delta \rho \right) +\rho _{o}\,\theta =0
\end{equation}
\begin{equation}
\rho _{o}\frac{\partial \theta }{\partial t}=-\,\frac{1}{\rho
_{o}\,\kappa _{T}}\delta ^{i\,j}\frac{\partial \left( \delta \rho
\right) }{\partial x^{i}\partial x^{j}}-\frac{\beta }{\kappa
_{T}}\delta ^{i\,j}\frac{\partial \left( \delta T\right)
}{\partial x^{i}\partial x^{j}}+D_{v}\delta ^{i\,j} \frac{\partial
\left( \theta \right) }{\partial x^{i}\partial x^{j}}-\delta
^{i\,j}\frac{\partial \left( \varphi \right) }{\partial
x^{i}\partial x^{j}} \label{cinco}
\end{equation}
\begin{equation}
\frac{\partial }{\partial t}\left( \delta T\right) =D_{T}\,\delta
^{i\,j} \frac{\partial \left( \delta T\right) }{\partial
x^{i}\partial x^{j}}-\frac{ \beta \,T_{o}}{\rho _{o}C_{p}\kappa
_{T}}\theta   \label{seis}
\end{equation}
assuming that $F_{k}=-\frac{\partial \,\varphi }{\partial x^{k}}$
. Also, $ D_{v}=(\frac{4}{3}\eta +\varsigma )\frac{1}{\rho _{o}}$.
Now, we notice that $\frac{\partial }{\partial t}\left( \delta
\rho \right) =-\rho _{o}\,\theta $ so we can reduce this set to
only two equations, Eq. (\ref{cinco}) and Eq. ( \ref{seis}).  We
now perform a few minor transformations introducing the speed of
sound of the fluid $C_{o}^{2}$ $=\left( \frac{\partial p}{\partial
\rho }\right) _{s}$ through the relationship
\begin{equation}
k_{T}=\gamma k_{s}=\frac{C_{p}}{C_{v}}k_{s}  \label{Sound1}
\end{equation}
where $k_{s}$ is the adiabatic compressibility $\left(
\frac{\partial \rho }{
\partial p}\right) _{s}$ whence, $k_{T}=\frac{\gamma }{C_{o}^{2}}$. This
leaves us finally with the set:
\begin{equation}
-\frac{\partial ^{2}(\delta \rho )}{\partial
t^{2}}+\frac{C_{o}^{2}}{\gamma } \delta ^{i\,j}\frac{\partial
\left( \delta \rho \right) }{\partial x^{i}\partial
x^{j}}+\frac{C_{o}^{2}\beta }{\gamma }\delta ^{i\,j}\frac{
\partial \left( \delta T\right) }{\partial x^{i}\partial x^{j}}+D_{v}\delta
^{i\,j}\frac{\partial }{\partial x^{i}\partial
x^{j}}(\frac{\partial }{
\partial t}\left( \delta \rho \right) )+\rho _{o}\frac{\partial \left(
\varphi \right) }{\partial x^{i}\partial x^{j}}=0  \label{sis1}
\end{equation}
\begin{equation}
\frac{\partial }{\partial t}\left( \delta T\right) =D_{T}\,\delta
^{i\,j} \frac{\partial \left( \delta T\right) }{\partial
x^{i}\partial x^{j}}-\frac{ \gamma -1}{\beta }\frac{\partial
}{\partial t}\left( \delta \rho \right) =0 \label{sis2}
\end{equation}
where we have used that $C_{p}-C_{v}=\frac{\beta ^{2}\,T_{o}}{\rho
_{o}^{2}\kappa _{T}}$.

\bigskip Eqs.(\ref{sis1}-\ref{sis2}) form a set of coupled equations for the
density and temperature fluctuations in the fluid under the action of a
conservative force whose nature need not to be specified for the time being.
They are the basis for studying the properties of the time correlation
functions of thermodynamic fluctuations. Those of the density will be of
particular interest here.

\section{Solution to the hydrodynamic equations}

The results derived in the previous section are far from being new. Aside
from the term $\nabla ^{2}\phi $ which arises from the presence of an
external conservative force, they are identical to the ones that have been
widely discussed in the literature. The question here is if the
gravitational potential introduces any substantial modification in the
correlation functions for the thermodynamic fluctuations. To examine this
possibility we recall that, if we are considering only the fluctuations, the
gravitational potential satisfies the Poisson equation:

\begin{equation}
\delta ^{i\,j}\frac{\partial \left( \varphi \right) }{\partial
x^{i}\partial x^{j}}=-4\pi G\,\delta \rho   \label{Sol1}
\end{equation}
Now, equation (\ref{sis2}) reads
\begin{equation}
-\frac{\partial ^{2}(\delta \rho )}{\partial
t^{2}}+\frac{C_{o}^{2}}{\gamma } \delta ^{i\,j}\frac{\partial
\left( \delta \rho \right) }{\partial x^{i}\partial
x^{j}}+\frac{C_{o}^{2}\beta }{\gamma }\delta ^{i\,j}\frac{
\partial \left( \delta T\right) }{\partial x^{i}\partial x^{j}}\
+D_{v}\,\delta ^{i\,j}\frac{\partial }{\partial x^{i}\partial
x^{j}}(\frac{
\partial }{\partial t}\left( \delta \rho \right) )-4\pi G\rho _{o}\,\delta
\rho =0  \label{sis3}
\end{equation}
The solution to Eqs.(\ref{sis2}) and (\ref{sis3}) proceeds in the
standard fashion. We reduce them to a set of algebraic equations
by taking their Laplace-Fourier transform, choose to set the
static temperature fluctuations equal to zero, eliminate the
temperature leading to an equation for $\delta \rho \left(
\vec{k},s\right) $, which is the ratio of two polynomials in $s$ .
To compute $\delta \rho \left( \vec{k},t\right) $ one must take
the inverse Laplace transform of the former quantity, which
demands the knowledge of the roots of the denominator, which is a
cubic equation in $ s^{3}$ (dispersion equation). Next, we \
proceed with intuition, we assume that the Rayleigh peak is nos
substantially modified by the external potential, so that
according to the standard theory, one of the roots of the cubic
equation is $ s=-D_{T}\,k^{2},$ where $D_{T}$ is the thermal
diffusivity. This allows us to simplify the denominator to the
form,
\begin{equation}
(s+D_{T}\,k^{2})(s^{2}+D_{v}k^{2}s+C_{o}^{2}k^{2}-4\pi G\rho _{o})=0
\label{dispnew}
\end{equation}
the two roots of the quadratic equation are:
\begin{equation}
s_{1,2}=-\frac{D_{v}k^{2}}{2}\pm i\left[ \left( C_{o}^{2}k^{2}-4\pi G\rho
_{o}\right) -\frac{D_{v}^{2}k^{4}}{4}\right] ^{1/2}  \label{rootsnew}
\end{equation}
Iff $4\pi G\rho _{o}=0$ and viscosity dominates over the term in $k^{2}$,
this result reduces to the one giving rise to the standard Brillouin peaks
which correspond to density fluctuations of the type \cite{GC}-\cite{Berne},
\begin{equation}
\delta \rho \left( \vec{k},t\right) =\delta \rho \left( \vec{k},0\right)
\frac{1}{\gamma }e^{-D_{v}k^{2}t}Cos\left[ C_{o}k\,t\right]
\label{Brillouin}
\end{equation}
which are the acoustic modes damped by the Stokes-Kirchhoff factor $D_{v}.$

\bigskip

On the other hand, if $4\pi G\rho _{o}\neq 0$, the threshold value
for $k$ distinguishing between damped oscillations and growing
modes is given by
\begin{equation}
\left( C_{o}^{2}k^{2}-4\pi G\rho _{o}\right) -\frac{D_{v}^{2}k^{4}}{4}=0
\label{Tnew}
\end{equation}
or,
\begin{equation}
k^{2}=\frac{2C_{o}^{2\,}}{D_{v}^{2}}\left( 1\pm \sqrt{1-\frac{4\pi G\rho
_{o}D_{v}^{2}}{C_{o}^{4\,}}}\right)   \label{threshold2}
\end{equation}
Eq. (\ref{threshold2}) is a generalization of Jeans wave number
when dissipative effects due to viscosity are non-negligible, and
is the main result of the paper. We can note that, if $\frac{4\pi
G\rho _{o}D_{v}^{2}}{ C_{o}^{4\,}}\ll 1,$ and taking the $-$ sign
for the square root, we have:
\begin{equation}
k^{2}\approx \frac{2C_{o}^{2\,}}{D_{v}^{2}}\left[ 1-1+\frac{1}{2}\left(
\frac{4\pi G\rho _{o}D_{v}^{2}}{C_{o}^{4\,}}\right) \right]   \label{limit1}
\end{equation}
or
\begin{equation}
k^{2}\approx \frac{4\pi G\rho _{o}}{C_{o}^{2\,}}=K_{J}^{2}  \label{limit2}
\end{equation}
This is the square value of the Jeans wave number \cite{Kolb}
\cite{Peebles}. It was derived by Jeans in 1902 and rederived in
many other waves by several authors \cite{Kolb}. Here we simply
show that it is almost a trivial consequence of LIT.

\bigskip

Let us expand the square root in Eq. (\ref{threshold2}) up to
second order in order to have a better grasp of the effects of
viscosity on the Jeans wave number. Now we have that,
\begin{equation}
k^{2}\approx
\frac{2C_{o}^{2\,}}{D_{v}^{2}}-\frac{2C_{o}^{2\,}}{D_{v}^{2}}
\left[ 1-\frac{2\pi G\rho
_{o}D_{v}^{2}}{C_{o}^{4\,}}-\frac{1}{8}\left( \frac{4\pi G\rho
_{o}D_{v}^{2}}{C_{o}^{4\,}}\right) ^{2}\right] \label{limit3}
\end{equation}
or, using Eq. (\ref{limit2}):
\begin{equation}
k^{2}\approx K_{J}^{2}+\frac{4D_{v}^{2}}{C_{o}^{6\,}}\left( \pi G\rho
_{o}\right) ^{2}=K_{J}^{2}(1+\frac{2D_{v}^{2}}{C_{o}^{4}}\pi G\rho _{o})
\label{limit4}
\end{equation}
so that the dimensionless number
\begin{equation}
\varepsilon =\frac{2D_{v}^{2}}{C_{o}^{4}}\pi G\rho
_{o}=\frac{2\,(\frac{4}{3} \eta +\varsigma )^{2}}{\rho
_{o}C_{o}^{4}}\pi G  \label{limit5}
\end{equation}
is a good measure of the\ second order corrections to the Jeans wave number
due to viscous effects.

\bigskip

For the corrected Jeans mass we have \cite{Kolb}:
\begin{equation}
M_{c}=\frac{4}{3}\pi \left( \frac{\pi }{k}\right)
^{3}=\frac{4}{3}\pi \left( \frac{\pi }{K_{J}^{\
}(1+\frac{2D_{v}^{2}}{C_{o}^{4}}\pi G\rho _{o})^{\frac{1
}{2}}}\right) ^{3}\rho
_{o}=M_{J}(1+\frac{2D_{v}^{2}}{C_{o}^{4}}\pi G\rho
_{o})^{-\frac{3}{2}}  \label{Mjeans1}
\end{equation}
or
\begin{equation}
M_{c}=M_{J}(1+\varepsilon )^{-\frac{3}{2}}  \label{Mjeans2}
\end{equation}
where the standard Jeans mass $M_{J}$ is given by\cite{Kolb}:
\begin{equation}
M_{J}=\frac{4}{3}\pi \left( \frac{\pi }{K_{J}^{\ }}\right)
^{3}\rho _{o}= \frac{4}{3}\pi \left( \frac{\pi }{\sqrt{4\pi G\rho
_{o}}}C_{o}\right) ^{3}\rho _{o}=\frac{\left( \pi \right)
^{\frac{5}{2}}}{6\left( \left( \rho _{o}\right)
^{\frac{1}{2}}\left( G\right) ^{\frac{3}{2}}\right) }C_{o}^{3}
\label{Mjeans3}
\end{equation}
One should compare these results with Weinberg \cite{Weinberg} and
Peebles p. 108 \cite{Peebles}. The question here is if in the
primeval Universe, the plasma had a substantial viscosity.

\bigskip

Eq. (\ref{Mjeans3}) deserves more attention. As shown by Weinberg
\cite {Weinberg} a similar result follows from a more intuitive
calculation addressing the behavior of the dispersion relation in
sound waves. Using numerical values for the density of barions
(protons) and the velocity of sound in a background of photons and
protons, he is able to estimate the temperature range for which
the protogalactic mass can grow to $10^{11}$ solar masses. Similar
reasoning may be applied to Eq. (\ref{Mjeans2}) and, of course,
predict a new $M_{J}$. The question here is if the viscous
corrections have any real meaning at all or they just turn out to
be an academic curiosity. For instance, in accretion disks
\cite{Bianchini} $\rho _{o}\approx 4\times
10^{-4}\frac{kg}{m^{3}}$, $\eta $ can be estimated as high as $
10^{10}Pa\cdot s\,$and $C_{o}\approx 5\times 10^{3}\frac{m}{s}$ .
This would yield a value of $D_{v}\approx 10^{14}\frac{m^{2}}{s},$
so that
\begin{equation}
\varepsilon =\frac{2D_{v}^{2}}{C_{o}^{4}}\pi G\rho _{o}\approx 10^{-1}
\label{acr1}
\end{equation}
the estimated correction is at the order of 10\%.

\bigskip

\section{Final Remarks}

Dissipation is always present in real processes. Whenever the
thermodynamical functions of a self-gravitating viscous system
fluctuate around an equilibrium value, dissipation will affect the
general equilibrium condition for the stability of those
fluctuations. We have been able to establish a simple expression,
Eq. (\ref{limit5}) that measures how important this correction
will be, assuming that the Rayleigh peak remains essentially
unaffected by the the gravitational field. If this were not the
case, then a full treatment of the dispersion relation would be
needed, involving interesting effects of the gravitational field
in the Rayleigh-Brillouin spectrum, for the case of damped
systems. This modifications may have some importance in
astrophysics, since natural astrophysical masers are already
well-known \cite{masers} and its scattering in fairly dense
systems will provide information about the transport properties of
the scatterer.

We thank Ana Laura Garc\'{\i}a-Perciante, and Roy Maartens for valuable
comments. AS-V was supported by a CONACyT postdoctoral grant, and thanks the
Relativity and Cosmology Group at Portsmouth for hosting him while this work
was done.

\end{document}